\documentstyle[12pt,psfig,aaspp]{article}
\begin{document}

\textwidth 6.5in
\textheight 8.5in
\topmargin 0.0in
\oddsidemargin 0.0in

\title{Lunar Outgassing, Transient Phenomena and The Return to The Moon\\
II: Predictions of Interaction between Outgassing and Regolith}

\author{Arlin P.S.~Crotts and Cameron Hummels}
\affil{Department of Astronomy, Columbia University,
Columbia Astrophysics Laboratory,\\
550 West 120th Street, New York, NY 10027}

\begin{abstract}
In this second paper in the series, we consider the implications from Paper I
on how gas leaking through the lunar surface might interact with the regolith,
and in what respects this might affect or cause the appearance of optical
Transient Lunar Phenomena (TLPs).
We consider briefly a range of phenomena, but concentrate at the extremes of
high and low gas flow rate, which might represent the more likely behaviors.
Extremely fast i.e., explosive, expulsion of gas from the surface is investigated
by examining the minimal amount of gas needed to displace a plug of regolith
above a site of gaseous overpressure at the regolith's base.
The area and timescale of this disturbance is consistent with observed TLPs.
Furthermore there are several ways in which such an explosion might be expected
to change the lunar surface appearance in a way consistent with many TLPs,
including production of obscuration, brightening and color changes.

At the slow end of the volatile flow range, gas seeping from the interior is
retained below the surface for extensive times due to the low diffusivity of
regolith material.
A special circumstance arises if the volatile flow contains water vapor,
because water is uniquely capable of freezing as it passes from the base to the
surface of the regolith.
For a large TLP site, it is plausible to think of areas on the km$^2$ scale
accumulating significant bodies of water ice.
Furthermore, as the system evolves over geological time, the ice accumulation
zone will evolve downwards into the regolith.
Many possible reactions between the volatiles and regolith can act to decrease
diffusivity in the regolith,  depending on water and additional gases.
Thus, it is plausible that the volatiles produce a barrier between the seepage
source and vacuum, forcing the ice zone to expand outwards to larger areas.

These effects will influence some of the techniques we consider to further
probe TLPs and other aspects of inreractions between outgassing and the
regolith.
Paper III is devoted to describing these proposed techniques.
\end{abstract}

\newpage
\section {Introduction}

From Paper I it is apparent that lunar outgassing and transient lunar phenomena
(TLPs) are correlated, and we are thereby motivated to understand if this
correlation is directly causal.
It is a major purpose of this paper to explore quantitatively how outgassing
might produce TLPs.
Furthermore, it seems likely that outgassing activity is concentrated in
several areas, which leads one to ask how outgassing might interact with and
alter the regolith presumably overlying the source of gas.
Some reviews of similar processes have been made, but few have been written
since the integration of Apollo-era data e.g., Stern (1999), Mukherjee
(1975), Friesen (1975) and none deal with the specific questions treated here.

It is evident from several experiments from the Apollo missions that gas is
being produced in the vicinity of the Moon, even though these experiments
differ on the total amount:
1) LACE (Lunar Atmosphere Composition Experiment on {\it Apollo 17}), $\sim$0.1
g s$^{-1}$ over the entire lunar surface (Hodges et al.~1973, 1974);
2) SIDE (Suprathermal Ion Detector Experiment on {\it Apollo 12, 14, 15}),
$\sim$7 g s$^{-1}$ (Vondrak et al.~1974), or $\sim$200 tonne y$^{-1}$;
3) CCGE (Cold Cathode Gauge Experiment on {\it Apollo 12, 14, 15}), $\la $60 g
s$^{-1}$ (Hodges et al.~1972).
These measurements not only vary by more than two orders of magnitude but in
assayed species and detection methods.
The LACE result here applies only to neutral $^{40}$Ar, $^{36}$Ar and $^{20}$Ne.
The SIDE results all involve solely ions, and perhaps include a large
contribution from molecular species (Vondrak et al.\ 1974).
CCGE measures all neutral species, and cannot easily distinguish between them.

The LACE data provides clear evidence for episodic outgassing of $^{40}$Ar on
timescales of a few months or less (Hodges \& Hoffman 1975), but the results
resolving this into faster timescales are more ambiguous.
For now we consider the average rate.
In the discussion below we will adopt the intermediate case, SIDE, for the
total production of gas, of all species, ionized or neutral.
The LACE is the only instrument to provide compositional ratios, which include
additional, rarer components.
We will use these ratios and in some cases normalize them against the SIDE
totals.

Many of the following discussions are only marginally sensitive to the actual
composition of the gas.
For many components of molecular gas at the lunar surface, there is a
possible contribution from cometary or meteoritic impacts, and a lesser amount
from the solar wind.
The influx of molecular gas from comets and meteorites is estimated usually in
the range of tonnes or tens of tonnes y$^{-1}$ over the lunar surface (see Anders
et al.\ 1973, Morgan \& Shemansky 1991).
Cometary contributions may be sporadically greater (Thomas 1974).
Except for H$_2$, solar wind interactions (Mukhergee 1975) provide only a small
fraction of the molecular concentration seen at the surface (which are only
marginally detected: Hoffman \& Hodges 1975).
There is still uncertainty as to what fraction of this gas is endogenous.
Current data do not succeed in resolving these questions, but we will return to
consider them later in the context of gas seepage/regolith interactions.

\section {Possible Modes of Outgassing/Regolith Interaction}

One can easily picture several modes in which outgassing volatiles might 
interact with regolith on the way to the surface.
These modes will come into play as one increases the flow rate of gas
(and/or varies the regolith overburden), and we simply list them with
mnemonic labels along with a description:

1) choke: complete blockage below the regolith, meaning that any chemistry or
phase changes occur within the bedrock/megaregolith;

2) seep: gas is introduced slowly into the regolith, on essentially a
molecule-by-molecule basis;

3) bubble: gas is introduced in macroscopic packets which stir or otherwise
rearrange the regolith (such as ``fluidization'' e.g., Mills 1969); 

4) gulp: gas is introduced in packets whose adiabatic expansion deposits
kinetic energy into regolith and cools the gas, perhaps even freezing it onto
regolith particles;

5) explode: gas is deposited in packets at base of the regolith leading to an
explosion; and

6) jet: there is little or no entrained material and gas simply flows into the
vacuum at the speed of sound.

While the intermediate processes might prove interesting, the extreme cases are
probably more likely to be in effect and will receive more of our attention.
In fact choking behavior might lead to explosions or jets, when the pressure
blockage is released.
Since these latter two processes involve primarily simple hydrodynamics (and
eventually, newtonian ballistics), we will consider them first, and how they
might relate to TLPs.

\section{Explosive Outgassing}

If outgassing occurs at a rate faster than simple percolation can sustain, and
at a location where regolith obstructs its path to the surface, the
accumulation of the gas will disrupt and cause bulk motion of the intervening
regolith.
The outgassing can lift the regolith into a cloud in the temporary atmosphere
caused by the event.
The presence of such a cloud has the potential to increase the local albedo
from the perspective of an outside observer due to increased reflectivity and
possible Mie scattering of underlying regolith.
Additionally, volatiles buried in the regolith layer might become entrained in
this gas further changing the reflective properties of such a cloud.

Let us construct a simple model of explosive outgassing at the lunar surface.
For such an event to occur, we assume a pocket of pressurized gas builds at the
base of the regolith, where it was delivered by transport through the
crust/megaregolith (presumably via channels/cracks or at least faster
diffusion).
Gas will accumulate at this depth until its internal pressure is sufficient to
displace the overlying regolith mass, or some release of downward pressure
occurs on a short timescale (impact, fluidization, puncturing a seal, seismic
disturbance, etc.)~
We can estimate the amount of gas alone required to cause explosive outgassing
by assuming that the internal energy of the buried gas is equal to the total
energy necessary to raise the overlying cone of regolith to the surface.
Of course, if there is any additional energy provided by the motion of delivery
of the gas itself, this will reduce the overall gas mass neccesary to displace
the overlying regolith.
This ``minimal TLP'' is the smallest outgassing event likely to produce
potentially observable disruption at a new site, although re-eruption through
thinned regolith will require less gas.

We can think of the outgassing event occurring in two parts.
Initially, the gas bubble explodes upward pushing the overlying regolith with
it until it reaches the level of the surface.
Through this process the gas and regolith become mixed.
At this point, the gas will expand and drag the regolith outward until it
reaches a sufficiently small density to allow the gas to escape freely into the
vacuum.
In the ``uncorking'' phase, we use a simple analytical model, whereas for the
expansion we numerically integrate averaged equations of motion in narrow
timesteps to follow the acceleration of the entrained regolith in the expanding
gas.

We discuss the depth of regolith further below; for this calculation we assume
a typical depth of 15 m.
Furthermore, we set the bulk density of the regolith at $1.9$ g cm$^{-3}$
(McKay et al.\ 1991), thereby setting the pressure at this depth at 0.45 atm. 
Because of the violent nature of an explosive outgassing, we assume that the
cone of dust displaced will be 45$^\circ$ from vertical (comparable to the
angle of repose for a disturbed slope of this depth: Carrier et al.\ 1991).
The mass of overlying regolith defined by this cone is $m= 7 \times 10^6$ kg.

By equating the energy necessary to lift the mass $m$ of the regolith cone $h =
15$ m ($U = mgh$, where $g = 1.62$ m s$^{-2}$) with the internal energy of the
buried gas, we require 47,000 moles of gas.
Much of this gas presumably consists of $^4$He, $^{36}$Ar and $^{40}$Ar (see
Paper I, op.\ cit.), so we assume a mean molecular mass for the model gas of
$\mu \approx 20$ AMU, hence 940~kg of gas is necessary to create an explosive
outgassing event.  
The temperature at this depth is $\sim0^\circ$C (see below), consequently
implying an overall volume of gas of 2300 m$^3$ or a sphere 8.2 m in radius.

We assume that this mass of gas forces the regolith plug out of the surface and
in the process the gas and regolith mix thoroughly in pressure, temperature,
density and composition.
After the plug is expelled, the gas
continues to expand in a hemisphere centered on the plug until it can escape
freely into the vacuum.
Free escape should occur when the mean free path of a gas particle (without
colliding with a dust particle) is on the order of the distance to the edge of
the expanding hemisphere.  
The behavior of this cloud is legislated by the evolution of pressure acting on
particles of varying cross-section.
For the purposes of this calculation we idealize the particles as spheres and
follow their evolution of motion in 512 logarithmically-spaced particle radius
bins consistent with lunar maria samples (McKay et al.\ 1991).

Following the evolution of this cloud in small, constant timesteps, we monitor
the typical trajectories of particles of different size.
During this expansion, the gas does $PdV$ work on the regolith, according to
the ideal gas law and partitioned among the particles according to their
cross section.
We assume that the gas's temperature stays in good thermal contact with the
regolith, making it nearly isothermal.
The total energy imparted to the dust is $W = NRT$ ln ($V_{final}/
V_{initial}) \approx 10^{9}$ J, equivalent to about one-quarter ton of TNT.
We let regolith particles fall from the distribution when the time since the
initial explosion exceeds a particle's ballistic ``hang time'' legislated by
its kinetic energy.
(We take the 45$^\circ$ case as typical.)

Because smaller particle sizes have larger cross-sectional area per unit mass and
are thus more effectively accelerated outward, the cloud quickly differentiates
with large particles dominating the central regions and small particles
predominantly on the periphery.
Figure 1 plots the hang times of different particles sizes versus time.
The regolith particles larger than 100 micron (about 50\% of the distribution
by mass) remain aloft for only a few seconds, a large fraction falling in or
near the explosion crater.
Particles smaller than optical wavelengths stay aloft for many minutes.

We integrate the area of regolith particles from the top of the cloud to
determine the optical thickness of different portions of the cloud.
Figure 2 shows the size evolution of the optically thick cloud ($\tau = 1$),
which quickly expands to 5 km in diameter, where it remains fixed until the
dust distribution is expended/collapses, about 1 minute after the explosion.
At optical thickness $\tau = 0.1$, which might indicate the limit of any effect
observable to the naked eye, the cloud expands quickly to 16 km diameter and
remains there for about 100 s.

Experiments with agitation of lunar regolith (Garlick et al.\ 1972) show that
the reflectance of dust is nearly always increased under fluidization,
typically by about 20\% and often by about 50\%, and similar results should be
expected here.
These increases in lunar surface brightness would be easily observable spread
over the many square kilometers indicated by the model.
Indeed, these areas and timescales would correspond to many reported TLPs at
the less striking end of the distribution.
Also, the cloud should cast a shadow that will be even more observable,
blocking the solar flux from a comparable area.
Furthermore, because the sub-micron particle sizes dominate the outer
regions of the cloud (the outer 1 km of the optically thick cloud), it seems
reasonable to expect Mie-scattering effects in this region with both blue and 
red clouds expected from different Sun-Earth-Moon orientations.

Some descriptions of TLPs, however, seem to indicate an increase in surface
brightness even greater than any described here.
In Appendix I, we discuss another mechanism which might both increase the
surface brightness and change the cloud's color, but drastic changes in these
quantities seem beyond the model presented above.

The success of this model in reproducing the basic size and timescale of many
TLP reports is encouraging, since it describes a basic process -- outgassing
forcing a puncture in the regolith layer.
How often such an event should take place is unknown, not only due to
uncertainty of the magnitude and distribution of endogenous gas flow to the
surface, but also with how the regolith will react to a surface puncture.
A new crater caused by explosive outgassing will cause the temperature 
structure of the regolith to adapt, with layers that are closer to the surface
after the explosion cooling to accomodate the lesser amount of insulation by
overlaying regolith.
We will not attempt here to follow the next steps in the evolution of an
outgassing ``vent'' in this way, but are inspired to understand how regolith,
its temperature profile and gas interact, as in the next section.

\section {Seepage of Gas Through the Regolith}

The question stands as to how fast the gas must accumulate in order to cause
the explosive event in the previous section, rather than simple seepage.
The escape of gas into the vacuum is regulated by the state of the regolith and
is presumably largely diffusive.

Of special importance is the measured abundance of small particle sizes evident
in the upper levels of the regolith, which pertains to depths $\sim
15$ m (where bulk density is probably higher: Carrier et al.\ 1991).
Assuming that particle distributions are self-similar in size distribution
(constant porosity), for random-walk diffusion out of a volume element $dV$,
the diffusion time step presumably scales with the particle size $a$, so the
diffusion time $t \propto a^{-1}$.
For particles of the same density, therefore, one should compute the diffusion
time by taking a $a^{-1}$-weighted average of particle sizes counted by mass.
$\langle a \rangle$.

This moment of the distribution is the same required in the previous section.
Published size distributions measured to sufficiently small sizes include again
McKay et al.~(1991) with $\langle a \rangle = 24$ $\mu$m, and supplemented on
smaller sizes with {\it Apollo 11} sample 10084 (Basu \& Molinaroli 2001),
which reduces the average to about 20 $\mu$m.
This is an overestimate because a large fraction (34-63\%) are agglutinates,
which are groupings of much smaller particles, many having effective areas e.g.,
$e = A/4 \pi r^2$, with values of a few up to 8.
(Here $r$ is a mean radius from the center of mass to a surface element.)~
To a gas particle, the sub-particle size is more relevant than the agglutinate
size, so the effective particle size of the entire sample might be considerably
smaller, conceivably by a factor of a few.
For the sake of discussion we adopt a random-walk step size $l$ of 10 $\mu$m.

At total gas pressure $P_g = 0.45$ atm in Section 3, the mean free path is of
order a few millimeters, much greater than $l$.
This is in the Knudsen flow regime, where the flux of gas is $J_k = 4 K \mu
(l / Z) P_g / \sqrt{2 \pi k T \mu}$ (e.g., Schorghofer \& Taylor 2007),
%J_k = 4.2e17(mu)/cm^2/s = 1.4e-5g/cm^2/s
where $K$ is a factor of order unity, $Z$ the depth below the surface (which
we presume is $\sim$15 m), $\mu$ the mean atomic mass (20 AMU adopted above),
and $T$ the temperature ($\sim$260K, see below).
This gives $J_k \approx 10^{-5}$ g cm$^{-2}$ s$^{-1}$ for these parameters, or
$\sim 20$ g s$^{-1}$ over the spatial scale of an explosive event like the one
considered above, several times larger than the SIDE outgassing rate.
Local episodic outgassing might exceed the SIDE rate.
To accumulate the amount of gas required to power one of these explosions
requires about 1~d of lunar gas output at this rate.
However, below, when we compare experimental simulation and Knudsen values, we
suspect that the actual diffusivity might be significantly lower, so the
required flow rate might also be lower (and the time to produce an event
longer).

Were it not for phase changes of venting gas within the regolith, the
composition of the gas might be a weak consideration in this paper (except
for perhaps the molecular/atomic mass).
Water plays a special role in this context (separate from others' concerns
regarding resource exploitation or astrobiology), in that it is the only
common substance encountering its triple point temperature in passing through
the regolith, at least in many locations.
This implies that even relatively small volatile flows containing water
might tend to freeze in place and remain until after the flow stops.
For water this occurs at 0.01$^\circ$C, corresponding to 0.006 atm in pressure
(the pressure dropping by a factor of 10 every $\sim$25$^\circ$.)~

Water is the only relevant substance to behave in this fashion, the next most
common substances may be large hydrocarbons such as nonane or benzene,
obviously not likely abundant endogenous effluents from the interior.
Also H$_2$SO$_4$ reaches its triple point, but changes radically with even
modest concentrations of water.
A similar statement can be made about HNO$_3$, which will not behave in its pure
state, either.
We consider only water's phase-change effects in this paper.

We compare water seepage to experimental simulations, a reasonably close analogy
being the sublimation of water ice buried up to 0.2 m below a medium of
simulant JSC Mars-1 (Allen et al.\ 1998) operating at $\sim 263^\circ$K and
7 mbar (Chevrier et al.~2007), close to lunar regolith conditions in temperature
and nearly at the water triple-point pressure (although we cannot be certain of
the pressure environment due to other species, which probably dominate).
Chevrier et al.~find their experimental situation corresponds to the lifetime of
800~y for a 1~m ice layer, 1~m below the surface.
The porosity of JSC Mars-1 is 44-54\%, depending on compactification, versus
$\sim$49\% for lunar soil at the surface and $\sim$40\% at a depth of 60~cm and
slightly lower at large depths (Carrier et al.\ 1991).
Lunar soil is slightly less diffusive even by solely this porosity measure.
The mean size $\langle a \rangle$ of JSC Mars-1 is 93 $\mu$m, $\ga$10 times
larger than that for {\it Apollo 17} and {\it 11} regolith, accounting for
agglutinates, so the corresponding timescale for lunar regolith material is,
very approximately, $\ga$10 ky (perhaps up to $\sim$30 ky).
Other simulants are more analogous to lunar regolith, so that such experiments
in the future might be made more relevant.

From the heat flow experiments at the {\it Apollo 15} and {\it 17} ALSEP sites,
(Langseth \& Keihm 1977), we know that just below the surface, the stable
regolith temperature is in the range of $247-253$K (dependent on latitude, of
course), with gradients (below 1-2 m) of $1.2-1.8$ deg m$^{-1}$, which
extrapolates to $0^\circ$C at depths of $13-16$ m below the surface.
This is too deep to be affected significantly by variations in heating over
monthly and lunar orbital precession timescales.
This is an interesting depth, since in many areas the regolith is not quite this
deep, as small as under a few meters near Lichtenberg (Schultz \& Spudis 1983)
and at the Surveyor 1 site near Flamsteed (Shoemaker \& Morris 1970) to depths at
Apollo sites (summarized by McKay et al.\ 1991) near the $0^\circ$C depths
calculated above, up to probably 20 m or more in the highlands, and 40~m north of
the south Pole-Aitken Basin (Bart \& Melosh 2005).
Presumably, the fractured zone supporting the regolith does not contain as many
small particles useful for retaining water ice, as we detail below, but may also
accumulate ice temporarily.
A more recent heat flow analysis (Saito et al.\ 2007) would place the $0^\circ$C
depth at least twice as far under the surface, increasing the lifetime of
retained volatiles, but for now we proceed with a more conventional,
shorter-lived analysis.

The current heat flow, $\sim 3 \times 10^{-6}$ W cm$^{-2}$ (Langseth et
al.~1972) would need to be $\sim 10$ times higher in order to place the 
$0^\circ$C within about a meter of the surface and make ice formation more
problematic.
(Note, the H$_2$O triple-point pressure corresponds to the regolith overburden
at 0.23 m, so ice at these shallow depths creates a dynamical instability.)~
In the case of the maria, this requires waiting until $\sim$3 Gy b.p., which is
probably sufficient for the Moon globally (see Spohn et al.\ 2001).
After this point the $0^\circ$C depth will recede into the regolith, while the
regolith layer is also growing.
Also, note that at this time the average surface temperature was cooler by
$\sim15^\circ$C due to standard solar evolution (Gough 1981 -- perhaps
$17^\circ$ lower in the highlands at 4 Gy b.p.).
Since the the thickness of regolith after 3 Gy b.p.\ grows at only about
1 m per Gy (Quaide \& Oberbeck 1975), within the maria the $0^\circ$C depth
sinks into bedrock/fractured zone.
Whatever interaction and modification might be involved between the regolith
and volatiles will proceed inwards, leaving previous epoch's effects between
the surface vacuum and the $0^\circ$C layer now at $\sim 15$ m.

We should ask more generally of the fate of subsurface ice accumulation,
perhaps from a time in the past when current flow measurements do not apply.
Converting a loss rate for 1 m below the surface to 15 m would involve the
depth ratio $R$.
%Farmer (1976) predicts an evaporation rate scaling as $R^{-1}$ (as opposed to
%the no-overburden analysis: Ingersoll 1970).
Experiments with varying depths of simulated regolith (Chevrier et al.\ 2007)
show that the variation in lifetime indeed goes roughly as $R^1$ as predicted
(Farmer 1976), implying a 1~m ice slab lifetime at 15 m on the order of $10^5$
to $3 \times 10^5$ y.
The vapor pressure for water ice drops a factor of 10 in passing from
$0^\circ$C to current temperatures of about $-23^\circ$C just below the surface
(also the naked-ice sublimation rate: Andreas 2007), which would indicate that
$\sim$90\% of water vapor tends to stick in overlying layers (without affecting
the lifetime of the original layer, coincidentally).
At the near-surface temperature at earlier times, 3-4 Gy b.p., this would be
98-99\%.
%Andreas 2007: 273K, 1e9 ug/cm^2/h; 247K, 1e8; 232K, 1.5e7; 230K, 1e7
It is worth speculating that at these epochs volatile production was presumably
higher, as would be ice production as soon as the temperature structure
allowed, but this is difficult to estimate.
At present, ice filling the interstices between regolith particles from 15 m to
1 m depth would amount to a surface density of $\sim 7$ tonne m$^{-2}$, hence it
would have a lifetime of order 1 My to a few My.

From above, there seem to be two evolution-dependent regimes that must be
considered in volatile-regolith seepage interactions.
The situation is entirely different when a gas source arises in pre-established
regolith, rather than one in which volatiles have been seeping over geological
time.
In the former case, we can simply use Fick's steady state diffusion equation
$\frac{\partial n} {\partial t}=-\kappa \nabla n$, where spatial gradients are
taken over 15~m, and scaling the JSC Mars-1 diffusivity of 1.7~cm$^2$ s$^{-1}$ to
0.17~cm$^2$ s$^{-1}$ for lunar regolith.
Since the triple-point pressure corresponds to number density $n = 2.4 \times
10^{17}$ cm$^{-3}$, $\partial n / \partial t = 2.7 \times 10^{13}$ s$^{-1}$,
so that the regolith water vapor atmosphere needs replenishment daily.
If now the total outgassing of 7 g s$^{-1}$ includes 0.1 g s$^{-1}$ of water,
this rate can maintain a total volume of 120 m$^3$ at the triple-point pressure.
How much ice this atmosphere can maintain depends on geometrical details of the
ice body, but it is probably of similar volume e.g., a few hundred m$^2$ at 1~m
thickness and typical porosity.

An alternative calculation is to consider the Knudsen flux given above, which
at the H$_2$0 triple-point conditions and a production rate of 0.1 g s$^{-1}$
implies that a subsurface slab of ice at 15 m depth can be maintained with a
topside surface area of 60 m$^2$.
However, the diffusivity from Chevrier et al.\ scaled to the same smaller
particle size for lunar regolith versus simulant JSC Mars-1 is $\sim 34$ times
smaller than the Knudsen value, hence it should support an ice slab of about
2000 m$^2$ in area.
There is a great deal of uncertainty in the input parameters, particularly the
H$_2$O input flux, but given this uncertainty it seems plausible that
subsurface ice slabs of local geographical significance might have formed.

In the longterm case the volatiles may have interacted with the regolith over
Gy, and due to the evolution of lunar heat flow versus regolith overburden,
that interaction will tend to progress downward in a layer between the gas
source and vacuum, thus potentially affecting the efficiency of volatile
retention.
This is primarily a question of whether minerals in solution precipitate in a
layer filling the interstices between regolith particles, or other reactions
accomplish a change in porosity.

There is little experimental work on the aqueous chemistry of lunar regolith
(which will vary due to spatial inhomogeneity).
Solution of lunar fines by water is greatly accelerated in the absence of
other gases such as O$_2$ and N$_2$ (Gammage \& Holmes 1975), and appears to
proceed by etching the numerous damage tracks from solar-wind particles.
This process acts to spread material from existing grains without reducing
their size (which would otherwise tend to increase porosity).
Liquid water is more effective than vapor, not surprisingly, and ice tends to
establish a pseudo-liquid layer on its surface.

This is a complex chemical system that will probably not be understood without
at least involving simulation experiments, but may involve prohibitive
timescales.
The major constituents are presumably silicates, which will migrate in
solution only over geologic time.
(On Earth, of order 30 My for quartz, 700 ky for orthoclase feldspar,
KAlSi$_3$O$_8$ and 80 ky for anorthite, CaAl$_2$Si$_2$O$_8$: Brantley 2004.)~
One might also expect the production of Ca(OH)$_2$, plus perhaps Mg(OH)$_2$
and Fe(OH)$_2$.
It is not clear that Fe(OH)$_2$ would oxidize to more insoluble FeO(OH), but
any free electrons would tend to encourage this.
It seems that the result would be generally alkaline.
Since feldspar appears to be a major component in some outgassing regions
e.g., Aristarchus (McEwen et al.\ 1994), one should also anticipate the
production of clays.
This is not accounting for water reactions with other volatiles e.g., ammonia,
which has been observed as a trace gas (Hoffman \& Hodges 1975) perhaps in
part endogenous to the Moon, and which near $0^\circ$C can dissolve in water
at nearly unit mass ratio (also to make an alkaline solution).
Carbon dioxide is a likely volatile constituent, and along with water can
metamorphose olivine/pyroxene into Mg$_3$Si$_4$O$_{10}$(OH)$_2$ i.e. talc,
albeit slowly under these conditions; in general the presence of CO$_2$ and
thereby H$_2$CO$_3$ opens a wide range of possible reactions into carbonates.
Likewise the presence of sulfur (or SO$_2$) opens many possibilities e.g.,
CaSO$_4 \cdot$H$_2$O (gypsum), etc.
Since we do not know the composition of outgassing volatiles in detail, we will
probably need to inform simulation experiments with further remote sensing or
in situ measurements.

The mechanical properties of this processed regolith is also difficult to 
predict.
Some of these hydroxides expand, but a likely effect might be filling of the
interstitial volume by material, which will raise its density and make it more
homogeneous.
It seems likely that any such void-filling will sharply reduce the rate of
diffusion.
Indeed many of the possible mineral combinations are cement-like, and
experiments with water and anorthositic lunar chemical simulants have produced
high quality cement without addition of other substances, except SiO$_2$
(Horiguchi et al.\ 1996, 1998).
Other substances that would be simple to produce are of very low hardness and
not of high ductility.
One needs to consider the effects of cracks or impacts into this medium; we
speculate that vapor or solution flow would tend to deliver ice and/or solute
to these areas and eventually act to isolate the system from the vacuum.
This is probably not a dominant process, since even the overturn timescale to
depths as shallow as 1~m is more than 1~Gy (Gault et al.\ 1974, Quaide \&
Oberbeck 1975)).
Craters 75~m in diameter will permanently excavate to a 15~m depth (e.g.,
Collins 2001, and ignoring the effects of crack and breccia formation), and are
formed at a rate of about 1 Gy$^{-1}$ km$^{-2}$ (extending Neukum et
al.\ [2001] with a Shoemaker number/size power-law index 2.9).
The large flow into unmodified regolith discussed above might be disrupted by
impacts roughly every 1~Gy.
An exceptional case is Aristarchus, which may have radically affected any
modifications resulting from outgassing in its vicinity, several hundred My
ago.

The geologically long-lived, modified-regolith situation is complex, but we
might hope to understand situations in which new outgassing vents have opened
and have not affected their environment radically.
We should consider a maximal outgassing site, presumably Aristarchus at about
60\% of the total (Paper I), and a perhaps more typical site, at a few percent
(say, 3\%) of the total lunar flow rate.
According to SIDE results, the total atmospheric production is 220 tonne/y.

How much of this is reasonable to consider as water?
We see that lunar atmospheric water production might be suppressed by the
regolith.
On the Earth, water is the predominant volcanic juvenile outgassing component
(Gerlach \& Graeber 1985, Rubey 1964).
While Earth is a radically different hydrological environment, we will see
that the amount of water being discussed below is $\la 10^{-6}$ the
compositional abundance on Earth.
Furthermore, the current water content in the atmosphere is much less than what
would affect hydration in lunar minerals (Mukherkjee \& Siscoe 1973 - although
some lunar minerals seem to involve water in their formation environments:
Williams \& Gibson 1972, Saal et al.\ 2007, and perhaps Akhmanova et al.\ 1979).

The current best limit on water abundance is from the sunrise terminator
abundances from LACE, which produces a number ratio of H$_2$O/$^{40}$Ar with a
central value of 0.014 (with $2\sigma$ limits of 0-0.04), which might indicate
an actual H$_2$O/$^{40}$Ar outgassing rate ratio up to 5 times higher, converting
spatial density to column density, given differing scale heights (Hoffman \&
Hodges 1975).
Adopting the LACE observation of $^{40}$Ar production of 0.04 g s$^{-1}$,
corresponds to 18 kg y$^{-1}$ for water (perhaps 5 times higher, 90 kg
y$^{-1}$).
Adopting the SIDE rate of 7 g s$^{-1}$ in the $\sim$20-44 AMU mass range, and
assuming most of these are $^{40}$Ar (Vondrak, Freeman \& Lindeman 1974: given
the much lower solar wind contributions of other species in this range), this
translates to 15 tonne y$^{-1}$ in water, in which case most of which must be
ionized upon escape into the vacuum.
The disagreement between SIDE and LACE is a major source of uncertainty
(perhaps due to the neutral/ionized component ambiguity).
We also need to address what fraction of seeping water vapor might be trapped
versus released in outgassing, which is now largely unknown.

In the case of initial eruption of an outgassing site, a prime site (presumably
Aristarchus) might be expected to produce a 10 tonnes per year in water, hence,
at a loss rate of a few tonne m$^{-2}$ My$^{-1}$ as above is capable of
maintaining a stable ice slab of order 1 km$^2$ or less in area.
For a minor site e.g., at 3\% of the total lunar outgassing rate, this is a
slab only of order 100 m or less in radius.
At sizes this small or smaller, one must consider the diffusion as a 
two-dimensional rather than 1-D problem, in which case maintaining an
equilibrium ice slab is even more problematic.
Of course, this entire discussion depends on the spatial source function for
the outgassing, which we have assumed is a point source but is currently
unknown.

This result assumes unmodified regolith.
Over geological time, in the case where the diffusion is sharply decreased by
volatile/regolith chemistry, the area of the ice patch will increase until the
boundary at the patch edge, where the regolith is unaltered by volatile
interaction, corresponds to the patch area just derived e.g., $\sim$1 km$^2$.
The current outgassing rate at a prime site and taking the depth of the regolith,
$\sim 15$ m, for this margin, this implies a long-term equilibrium patch of order
300 km$^2$, or $\sim 10^9$ tonne of ice at 3 tonne m$^{-2}$.
At the current rate this mass might require several hundred My to accumulate,
given all of these assumptions.
Presumably the regolith at this margin might still become modified, and it is
unlikely that outgassing has remained constant over the past 3 Gy, but we will
not explore these possibilities further here.

Also recall that Aristarchus, along with Kepler, Copernicus and Tycho compose
60-70\% of the robust events in Paper I, and all of these are major impact
craters whose floors are likely to be covered with impact breccia and
solidified impact melt, both likely to be largely impermeable except for
surface cracks.
Deeper cracks caused by these recent, major impacts may channel gas to these
areas and may set up further regions relatively disconnected from the vacuum,
hence they may be places of interest in terms of volatile collection.

The conclusion that we draw from the previous discussion is that it is not
unreasonable to suspect that water ice might accumulate in large masses in the
vicinity of major outgassing sites, but that these would cover relatively
small areas of the Moon, perhaps at the very most hundreds or thousands of
km$^2$, not even 0.01\% of the lunar surface (total of $3.8\times 10^7$ km$^2$).
In some way the current analysis evokes early hypotheses of subsurface ice
(Gold 1962), but on a radically smaller scale.
These sites are likely limited to a very small number of small areas where most
of the outgassing occurs, not spread throughout the lunar surface.

The conclusion to be drawn, we reason, is that while subsurface water ice
accumulation at outgassing sites is unlikely to be globally significant or even
detectable in the mean properties of lunar materials, it may be a telltale tracer
in particularly active areas to indicate the role of water as a minor constituent.
With our current state of knowledge we might guess that the Aristarchus region
is the first and perhaps only region worth investigating.
Any accumulated ice will be covert, being unable to exist at the surface, but
might conceivably be spread over areas under the surface that are nonetheless
amenable to detection by remote sensing.
It is unclear if our current samples of lunar materials are relevant to this
assertion, in that none have been returned from the regions most suspected of
outgassing, and the total amount of hydration produced by these effects is
negligible when averaged over the lunar surface.
In Paper III we outline methods which might be employed to resolve this issue.

%1tonne/y over 100000y <-> 14000 m^2 @ 7tonne/m^2 => over 3Gy, 3km^3

\section {Discussion and Conclusions}

We are sensitive to the controversial nature of suggesting small but significant
patches of subsurface water ice, given the history of the topic.
In this series of papers, we take care to avoid ``cargo cult science'' --
selection of data and interpretion to produce dramatic but subjectively biased
conclusions that do not withstand further objective scrutiny (Feynman 1974).
In Paper III we offer several straightforward and prompt tests of our conclusions
and speculations which offer the prospect of settling many of these issues,
although the intrinsically hidden nature of ice might make resolution difficult.
We are operating in many cases in a regime where interesting observations have
been made but the parameters e.g., the endogenous lunar molecular production
(water vapor or otherwise), required to evaluate alternative models and
interpretations are sufficiently uncertain so as to frustrate immediate progress.
Despite the advances made primarily by Apollo-era research, we are still
skirting the frontiers of ignorance.

A case in point is the interpretation of outbursts of molecular gas near the
lunar surface.
At least two instances of these were published with assertions of
non-anthropogenic origin (Freeman et al.~1973, Hodges et al.~1973, 1974)
and another has been discussed without publication (Criswell \& Freeman 1975).
In the former two cases, discussions decades later (published and unpublished)
have dismissed these case as anthropogenic despite the lack of new data
(Freeman \& Hills 1991, Hodges 1991).
Initially the many SIDE outgassing events (Vondrak et al.\ 1974) were
categorized neither as anthropogenic, endogenous or cometary/meteoroid
impact-generated; these events' nature have not been treated explicitly in
publication.
It would be useful for further lunar volatile analyses if a review of their
probable origin were in hand; this is beyond the goal of this paper.
Likewise an understanding of the large differences in fluxes seen by SIDE, LACE
and the cold-cathode particle detectors would aid our analysis (see above).

Our results in this paper and in Paper I bear directly on the argument of
Vondrak (1977) that TLPs as outgassing events are inconsistent with SIDE
episodic outgassing results.
The detection limits from ALSEP sites {\it Apollo 12, 14} and {\it 15}
correspond to 16-71 tonne of gas per event at common TLP sites, particularly
Aristarchus.
(Vondrak states that given the uncertainties in gas transportation, these
levels are uncertain at the level of an order of magnitude.)~
Our ``minimal TLP event'' described above is 20-100 times smaller than this,
however, and still visible from Earth.
It seems implausible that a spectrum of such events would never exceed the SIDE
limit, but it is not so obvious such a large event would occur in the
seven-year ALSEP operations interval.

The other impact of our work on the SIDE analysis is the placement of robust
TLP sites relative to ALSEP sites.
Most notably Alphonsus, adjacent to Apollo landings (particularly 14 and 12),
drops from our list of persistent sites.
We cannot be sure at present if Alphonsus undergoes dormant periods or simply
is not a site of verifiable reports, but the SIDE upper limit of only 6.5 tonne
per event for Alphonsus is no longer a meaningful constraint on TLPs.

The only persistent site of some statistical significance that corresponds to a
sample return (170 g from Luna 24) is Mare Crisium, which is ambiguous in
several ways: 1) the robust TLP report count for Mare Crisium is only zero to 6,
depending on the robustness filter employed (Paper I);
2) the nature of this robust count is problematic given the extended nature of
Mare Crisium as a feature, and
3) the sample returned from Mare Crisium is one case where the absence of water
is ambiguous (Akhmanova et al.\ 1979), making volatile seepage in the vicinity
harder to dismiss.
In contrast, for Aristarchus the nearest sample return came from about 1100 km
away ({\it Apollo 12}).

This is not to say that lunar samples have never produced clear evidence of
endogenous water; picritic glass (Saal et al.\ 2007) fairly convincingly
contains
H$_2$O at the level of 4-46 ppm ($\pm$2 ppm), in a way that points to magmatic
origin.
Previous studies of other lunar minerals e.g., apatite, have likely overlooked
such contributions (McCubbin et al.\ 2007), whereas particular samples
occasionally have produced evidence of H$_2$O and other volatiles
(Kr\"ahenb\"uhl et al.\ 1973).
The statement is often made in lunar geology that the Moon is ``waterless,''
``totally dry,'' or the semantic equivalent.
While it is clear on the basis of available evidence that such statements are
true in spirit; formally or literally they are almost certainly incorrect.
The question is the amount of uncertainty regarding the abundance of water
within the Moon, either in terms of a tolerance or probability.

While this seems like a trivial point, it is not.
While the temperature of the proto-Earth and progenitor impactor material in
simulations grow to thousands of Kelvins, sufficient to drive off the great
majority of all volatiles, these are not necessarily the only masses in the
system.
Either body might have been orbited by satellites containing appreciable
volatiles, which would likely not be heated to a great degree and which would
have a significant probability of being incorporated into the final Moon.
Furthermore, there is recent discussion of significant water being delivered to
Earth/Moon distances from the Sun in the minerals themselves (Lunine et
al.\ 2007, Drake \& Stimpfl 2007), and these remaining mineral-bound even at
high temperatures up to 1000K (Stimpfl et al.\ 2007).
The volume of surface water on Earth is at least $1.4 \times 10^9$ km$^3$, so
even if the specific abundance of lunar water is depleted at the 99.9999\%
level, one should still expect tens of billions of tonnes endogenous to the
Moon, and it is unclear that later differentiation would eliminate this.
This residual quantity of water would be more than sufficient to concern us
with the regolith seepage processes outlined above.

Regardless of what one might believe on the controversial topics of endogenous
water or the non-radiogenic composition of lunar outgassing, it seems clear
that a simple model for the effects of gas from the interior of the Moon trying
to penetrate the regolith might easily be consistent with the timescales and
areas often quoted in TLP reports.
There is a plausible mechanism to explain at least the magnitude of the
phenomenon.
Explaining every reported detail of TLPs is beyond the scope of this paper, and
undoubtedly involves complex processes.
While we deal with one of these in Appendix I, we think it probably best at
this point to amass an objectively observed, relatively unbiased sample of
lunar optical transients (Crotts et al.\ 2007).
Providing the database from such a survey should be our goal given the much
improved state of available technology.

%For instance, the expected final H$_2$O ice fraction in a 350 km-diameter
%Main-belt planetesimal can be as high as 5\% in thermal history simulations
%and may be much larger than this to account for the density of Ceres (Farrell
%et al.\ 2007).

\bigskip
\bigskip

\section {Appendix: Coronal Discharge Luminescence Caused by Outgassing}

In Section 3 we see that it is possible for dust particles to remain suspended
for significant time intervals in a gas of number density $10^{11}$ to
$10^{15}$ cm$^{-3}$ over scales of a several tenths to several km; this seems a
plausible situation for generation of significant electrostatic potentials.
Considering that the regolith is composed of several predominant mineral 
compositions in various particle size ranges, under the action of suspension
and acceleration by gas flow, dust particle charge segregation is possible.

For the following calculation we adopt a typical particle size of
$r = 10$ $\mu$m, and a typical difference in work function $\Delta W = 0.5$ eV.
The actual value $\Delta W$ for particles of even well-defined compositions is
problematic due to various surface effects such as solar-wind/micrometeoritic
weathering and exposed surface Fe$^{2+}$ states.
The following analysis suffices for two particles of different conducting
composition; a similar result arises via triboelectric interaction of two
different dielectrics (although the details are less well-understood).
Disturbed dust is readily charged for long periods in the lunar surface
environment (Stubbs, Vondrak \& Farell 2005).
 
Two particles will exchange charge upon contact until the equivalent of
$\pm$0.25V is maintained, which amounts to $Q = CV = 4 \pi \epsilon_0 r V =
2.8 \times 10^{-16}$ coulomb = 1700 e$^-$.
%C = 0.0022 pF  $\epsilon_0 =8.8542\times 10^{-12}$ F m$^{-1}$
%F m$^{-1}$ = C$^2$ N$^{-1}$ m$^{-2}$
When these two particles are drawn apart to a distance $d >> r$ their mutual
capacitance becomes $C_2 = 4 \pi \epsilon_0 r^2/d$.
For $d = 100$~m, assuming $Q$ remains with the two particles, the voltage
increases by 7 orders of magnitude! - which cannot be maintained.
Note that Paschen's curve for coronal discharge goes through the minimal
potential at 137V for argon, 156V for helium, at column densities $N$ of
$3.2 \times 10^{16}$ cm$^{-2}$ and $1.4\times 10^{17}$ cm$^{-2}$, respectively.
In carefully controlled conditions, breakdown in He at potentials as low as 20V
can be achieved (Compton, Lilly \& Olmstead 1920).
Nominally, however, Paschen's curves usually rise steeply for lesser column
densities, and roughly proportional to $N$ for larger column densities.

We take He and Ar as the primary outgassed components, since $^{20}$Ne is less
well-established (Stern 1999) and likely of predominantly solar-wind origin
(Hodges et al.\ 1974), and the abundance of molecular gas is also uncertain.
The minimal-discharge column densities for molecules are similar to those for Ar
and He, and minimum voltages are several time higher e.g., 420V at $1.8 \times
10^{16}$ cm$^{-2}$ for CO$_2$; $\sim$430V for NH$_3$, 414V for H$_2$S, and 410V
for CH$_4$, and similar $N$ values for molecules e.g., $2.1 \times 10^{16}$
cm$^{-2}$.
These results are somewhat dependent of the structure and composition of the
electrodes used in making these measurements.

The visual appearance of atomic emission of these gases in high voltage discharge
tubes is well know, with He emitting a pink-orange glow (primarily from
transitions at 4471.5\AA, and 5875.7\AA, plus 7065.2\AA\ marginally visible to
the naked eye: Reader \& Corliss 1980), Ar a violet glow (from many lines
4159\AA\ to 4880\AA), and Ne, if present, producing intense red emission (with
many lines 5852\AA\ to 7032\AA).
It is plausible that TLP emission reaching coronal discharge conditions would
produce similar output spectra; the incidence of intense red emission in some
TLP reports (Cameron 1978) argues for the presence of Ne; some other reports
seem consistent with He and a smaller portion with Ar.
Given the non-endogenous nature inferred for most Ne, however, we need to
search for another gas in the case of very red events.
The most common candidate molecules produce coronal emission that appears
white (CO$_2$) or red (water vapor - primarily H$\alpha$, which is produced in
many hydrogen compounds; CH$_4$ - Balmer lines plus CH bands at 390 and 431 nm).

In comparison, the events and their colors actually reported can be counted,
although there is a significant range in outcomes depending on how one
categorizes mixed colors and other factors.
Following Paper I, we divide the sample at year 1956 (the latter composing 1/4
of the sample); before 1956: red - 11, blue (or blue-green) - 7, violet - 5,
yellow - 4, red-yellow - 3, brown - 2, orange - 2;
after 1955: red - 52, violet - 3, orange - 1.
In the later reports, three red events also contain elements of violet/blue,
plus one with orange, plus one with blue.
Evidently red reports are common (out of the total sample of 894, not including
the reports by Bartlett, which are largely blue), however, there is a
statistically significant change after 1956, where about 1/5 of reports include
red color, many associated with Aristarchus.
This change is presumably due to a shift in the nature of observer reporting
after 1955 rather than a change in the physics of the lunar events.

%yellow	violet	red		reddish	golden	blue(-	orange	brown
%<1956				-yellow	-brown	green)
%||||	||||\	||||\||||\|	|||	|	||||\||	||	|
%>1955
%	|||	52*					|
%*3 followed by violet/blue, 1 also orangish, 1 also blue

The initial gas density at the surface from a minimal TLP is of the order
$10^{18}$ cm$^{-3}$, meaning that the optimal column for coronal discharge
might conceivably be acheived on centimeter scales, whereas the initial
outburst is over tens of meters.
By the time a minimal TLP has expanded to 1 km radius, the density has dropped
to $\sim 10^{13}$ cm$^{-1}$, so the Paschen minimal $N$ holds over roughly the
scale of the entire cloud, which is likely the most favorable condition for
initiation of coronal discharge.
If gas kinetic energy is converted to luminescence with, for instance, 2\%
efficiency, at this density this amounts to $\sim 0.1$ J m$^{-3}$, or
100 J m$^{-2}$, compared to the Solar Constant (1366 W m$^{-2}$, with a typical
albedo of 7\%, yielding 100 W m$^{-2}$), so might be visible as a color shift
for at least several seconds.
This argues that a prolonged discharge luminescence effect, including color
changes, would require a larger, sustained event (more than the minimal $\sim$1
tonne), since the lifetime of some colored TLPs reports is many minutes.
It is not reasonable to think that a minimal TLP would sustain a coronal
discharge over its entire $\sim 45$ s lifetime sufficient to produce a visible
color change.

%minimal TLP: 45deg, 15m deep => 3500m^3; 938kg, 20AMU => 47000 mole= 2.8e28
%initial # density ~ 8e18cm^-3; dust radius= 816m => # density = 2.5e13cm^-3

Unless, in initiating the discharge, solar photoionization is important (which
seems unnecessary), these phenomena should also be observable on the nightside
surface, too, of which at least 20 are reported, usually a bright and/or variable
spot, 8 at/near Aristarchus e.g., 1824 May 1, near Aristarchus ``blinking light,
9th to 10th mag on dark side; 1881 February 3, near Aristarchus: ``very bright,
like an 8th mag star, pulsating;'' 1789 May 29: ``flickering spot on east edge of
Grimaldi,''  etc.
A rough calculation of Earthshine lunar surface brightness gives $\sim 12$ to
13 mag arcsec$^{-2}$ in V, compared to 3.4 mag for full sunlight.

The kernel of many of these ideas has been suggested earlier (Mills 1970), and
there is little doubt that luminous electrical discharges are generated in
terrestrial volcanic dust clouds (Anderson et al.\ 1965, Thomas et al.\ 2007).
In general reddish discharges may indicate H$\alpha$ from dissociation of a
number of molecular species, although is not the only possible explanation.
Visual descriptions alone are probably not sufficient to settle this question,
and the few spectra obtained of reported TLPs are sufficiently indistinct e.g.,
the Kozyrev (1958) spectrum, reported to show a transient spectrum of C$_2$
Swan bands.
Note that this observation was among those eliminated statistically by our
robustness filters in Paper I.
Kozyrev (1963) also reported transient H$_2$ emission from Aristarchus
(apparently absent Balmer lines).

Given the paucity of information from existing spectra and visual color
observations, we advance in Paper III efficient means to obtain further spectra
of TLP phenomena given a reliable imaging detection monitor, for which we are
operating a prototype.
Given the suggestion that red events might indicate one of several molecules,
it is worth discussing what wavelength range might be best to observe for
confirmation.
Rather than relying on what might be a bright H$\alpha$ line, plus faint
optical lines and bands required to distinguish molecules e.g., CH$_4$ versus
H$_2$O versus NH$_3$, we note that vibrational rotational bands for these (and
other molecules) are brighter and more discriminatory in the near-infrared.
K-band spectroscopy is likely to produce bright, distinctive features for these
and many other molecular candidates, even if only H$\alpha$ might be the only
optical transition readily detected.
This is discussed further in Paper III.

\noindent
\section{References}

\noindent
Akhmanova, M.V, Dement\'yev, B.V.\ \& Markov, M.N.\ 1979, Geochem.\ Internat.,
15, 166.

\noindent
Allen, C.C., et al.\ 1998, LPSC, 29, 1690.

\noindent
Anders, E. Ganapathy, R., Kr\"ahenb\"uhl, U.\ \& Morgan, J.W.\ 1973, The Moon,
8, 3.

\noindent
Anderson, R., et al.\ 1965, Science, 148, 1179.

\noindent
Andreas, E.L.\ 2007, Icarus, 186, 24.

\noindent
Bart, G.D.\ \& Melosh, H.J.\ 2005, D.P.S., 57, 57.07.

\noindent
Basu, A.\ \& Molinaroli, E.\ 2001, Earth, Moon \& Plan., 85, 25.

%\noindent
%Carrier, W.D.\ 1973, Moon, 6, 250.

\noindent
Brantley, S.L.~2004, in ``Treatise on Geochemistry'' eds.~H.D.~Holland \&
K.K.~Turekian (Elsevier: Amsterdam), section 5.03.

\noindent
Carrier, W.D., Olhoeft, G.R.\ \& Mendell, W.\ 1991, in ``Lunar Sourcebook,''
eds.~G.H.\ Heiken, D.T.\ Vaniman \& B.M.\ French (Cambridge U.: Cambridge),
p.\ 475.

\noindent
Chevrier, V., et al.\ 2007, GRL, 34, L02203.

\noindent
Collins, G.S.\ 2001, LPSC, 32, 1752.

\noindent
Compton, K.T., Lilly, E.G.\ \& Olmstead, P.S.\ 1920, Phys.\ Rev., 16, 282.

\noindent
Criswell, D.R.\ \& Freeman, J.W., Jr.\ \& 1975, The Moon, 14, 3.

\noindent
Crotts, A.P.S.\ 2007a, Icarus, submitted (PAPER I).

\noindent
Crotts, A.P.S.\ 2007b, ApJ, submitted (PAPER III).

\noindent
Drake, M.J.\ \& Stimpfl, M.\ 2007, LPSC, 38, 1179.	

\noindent
Farmer, C.B.\ 1976, Icarus, 28, 279.

\noindent
Farrell, L.L., McGary, R.S.\ \& Sparks, D.W.\ 2007, LPSC, 38, 1827.

\noindent
Feynman, R.P.~1974, Engineering \& Science, 37, 7.

\noindent
Freeman, J.W.\ \& Hills, H.K.\ 1991, Geophys.\ Res.\ Let., 18, 2109.

\noindent
Freeman, J.W., Hills, H.K.\ \& Vondrak, R.R.\ 1973, Proc.\ Lunar Sci.\ Conf., 3, 2217.

\noindent
Friesen, L.J.\ 1975, Moon, 13, 425

\noindent
Gammage, R.B.\ \& Holmes, H.F.\ 1975, LPSC, 6, 3305.

\noindent
Garlick, G.F.J., Steigmann, G.A., Lamb, W.E.\ \& Geake, J.E.\ 1972,
Proc.\ Lunar Sci.\ Conf., 3, 2681.

\noindent
Gault, D.E., H\"orz, F, Brownlee, D.E.\ \& Hartung, J.B.\ 1974,
Abs.\ Lun.\ Plan.\ Sci.\ Conf., 5, 260.

\noindent
Gerlach, T.M.\ \& Graeber, E.J.\ 1985, Nature, 313, 274

\noindent
Gold, T.\ 1962, in ``The Moon'' (IAU Symp.\ 14), eds.\ Z.\ Kopal \&
Z.K.\ Mikhailov (Academic: New York), p.\ 433.

\noindent
Gough, D.O.\ 1981, Solar Physics, 74, 21.

\noindent
Hodges, R.R.\ 1991, {\it personal communication}, in Stern, A.\ 1999, 
Rev.\ Geophys., 37, 4.

\noindent
Hodges, R.R., Jr., Hoffman, J.H., Yeh, T.T.J.\ \& Chang, G.K.\ 1972, JGR, 77,
4079

\noindent
Hodges, R.R., Jr., Hoffman, J.H., Johnson, F.S.\ \& Evans, D.E.\ 1973,
LPSC, 4, 2855.

\noindent
Hodges, R.R., Jr., Hoffman, J.H.\ \& Johnson, F.S.\ 1974, Icarus, 21, 415.

\noindent
Hodges, R.R., Jr.\ \& Hoffman, J.H.\ 1975, LPSC, 6, 3039.

\noindent
Hoffman, J.H.\ \& Hodges, R.R., Jr.\ 1975, Moon, 14, 159.

\noindent
Horiguchi, T., Saeki, N., Yoneda, T., Hoshi, T.\ \& Lin, T.D.\ 1996, in ``Space
V: 5th Internat'l Conf.\ Engin., Constr.\ \& Operat.\ in Space,''
ed.\ S.W.\ Stewart, ASCE Proc., 207, 86.

\noindent
Horiguchi, T., Saeki, N., Yoneda, T., Hoshi, T.\ \& Lin, T.D.\ 1998, in ``Space
`98: 6th Internat'l Conf.\ Engin., Constr.\ \& Operat.\ in Space,''
eds.\ R.G.\ Galloway \& S.L.\ Lokaj, ASCE Proc., 206, 65.

\noindent
Ingersoll, A.P.\ 1970, Science, 168, 972.

\noindent
Kozyrev, N.A.\ 1958, Sov.\ Intern't'l Geophys.\ Yr.\ Bull., PB 13162-42 (see
also 1962, in {\it The Moon, IAU Symp.\ 14}, eds.\ Z.\ Kopal \& Z.K.\ Mikhailov
(Academic: ), p.\ 263.

\noindent
Kozyrev, N.A.\ 1963, Nature, 198, 979.

\noindent
Kr\"ahenb\"uhl, U., Ganapathy, R., Morgan, J.W.\ \& Anders, E.\ 1973, Science,
180, 858.

\noindent
Langseth, M.G., Clark, S.P., Chute, J.L., Keihm S.J.\ \& Wechsler, A.E.\ 1972,
Moon, 4, 390.

\noindent
Langseth, M.G.\ \& Keihm S.J.\ 1977, in {\it Soviet-American Conference on
Geochemistry of the Moon and Planets} (NASA SP-370), pp. 283.

\noindent
Lunine, J., Graps, A., O'Brien, D.P., Morbidelli, A., Leshin, L.\ \& Coradini,
A.\ 2007, LPSC, 38, 1616.

\noindent
McCubbin, F.M., Nekvasil, H.\ \& Lindsley, D.H.\ 2007, LPSC, 38, 1354.

\noindent
McEwen, A.S., Robinson, M.S., Eliason, E.M., Lucey, P.G., Duxbury, T.C.\ \&
Spudis, P.D.\ \& 1994, Science, 266, 1858.

\noindent
McKay, D.S., et al.\ 1991, in ``Lunar Sourcebook,'' eds.\ G.H.\ Heiken,
D.T.\ Vaniman \& B.M.\ French (Cambridge U.: Cambridge), p.~285.
%chap. 7; also Langseth \& Keihm 1977; Quade \& Oberbeck 1975, ch. 4

\noindent
Mills, A.A.\ 1969, Nature, 224, 863

\noindent
Mills, A.A.\ 1970, Nature, 225, 929.

\noindent
Morgan, T.H.\ \& Shemansky, D.E.\ 1991, JGR, 96, 1351.

\noindent
Mukherjee, N.R.\ 1975, The Moon, 14, 169.

\noindent
Mukherjee, N.R.\ \& Siscoe, G.L.\ 1973, JGR, 78, 1741.

\noindent
Neukum, G., Ivanov, B.A.\ \& Hartmann, W.K.\ 2001, Space Sci.\ Rev., 96, 55.

\noindent
Reader, J.\ \& Corliss, C.H.\ 1980, CRC Handbook of Chemistry and Physics, 68.

\noindent
Quaide, W.\ \& Oberbeck, V.R.\ 1975, The Moon, 13, 27.

\noindent
Rubey, W.W.\ 1964, in ``Origin \& Evolution of Atmospheres \& Oceans,''
eds.\ P.J.~Brancazio \& A.G.W.~Cameron (Wiley: New York), p.~1.

\noindent
Saal, A., Hauri, E.H., Rutherford, M.J.\ \& Cooper, R.F.\ 2007, LPSC, 38, 2148.

\noindent
Shoemaker, E.M.\ \& Morris, E.C.\ 1970, Radio Sci., 5, 129.

\noindent
Schultz, P.H.\ \& Spudis, P.D.\ 1983, Nature, 302, 233.

\noindent
Schorghofer, N.\ \& Taylor, G.J.\ 2007, JGR, 112, E02010.

\noindent
Saito, Y., Tanaka, S., Takita, J., Horai, K.\ \& Hagermann, A.\ 2007, LPSC, 38,
2197.

\noindent
Spohn, T, Konrad, W., Breuer, D.\ \& Ziethe, R.\ 2001, Icarus, 149, 54.

\noindent
Stern, A.\ 1999, Rev.\ Geophys., 37, 4.

\noindent
Stimpfl, M., de Leeuw, N.H., Drake, M.J.\ \& Deymier, P.\ 2007, LPSC, 38, 1183.	

\noindent
Stubbs, V.J., Vondrak, R.R.\ \& Farrell, W.H.\ 2005, LPSC, 26, 1899.

\noindent
Thomas, G.E.\ 1974, Science, 183, 1197.

\noindent
Thomas, R.J, et al.\ 2007, Science, 315, 1097.

\noindent
Vondrak, R.R.\ 1977, Phys.\ Earth Plan.\ Interiors, 14, 293.

\noindent
Vondrak, R.R., Freeman, J.W.\ \& Lindeman, R.A.\ 1974, LPSC, 5, 2945

\noindent
Williams, R.J.\ \& Gibson, E.K.\ 1972, Earth Plan.\ Sci.\ Let., 17, 84.

\begin{figure}
\plotone{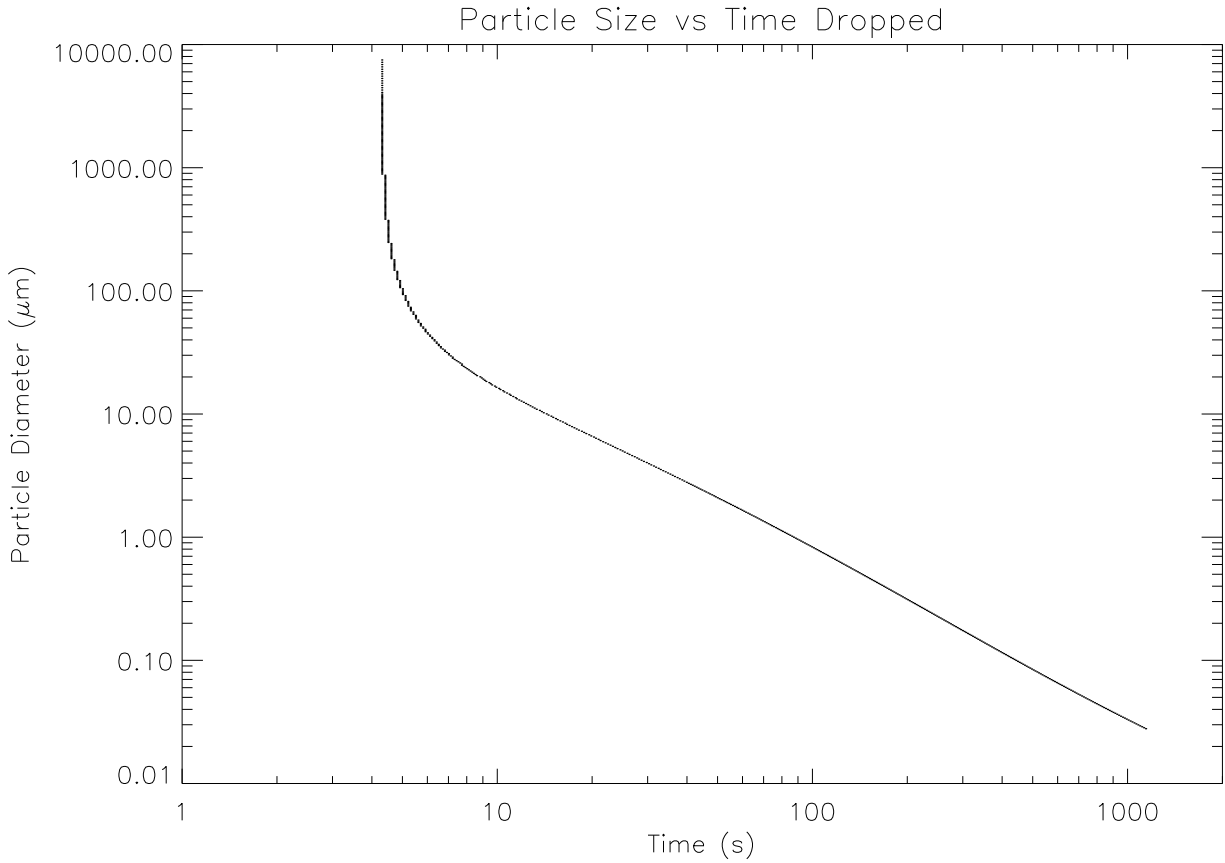}
\caption{
The time required for a typical regolith particle of the diameter shown to
fall out of the expanding cloud back to the lunar surface.
Particles larger than about 100 micron (about 50\% by mass) rain immediately to
the ground, barely exiting the initial crater, whereas particles smaller than
optical wavelength remain aloft for at least several minutes.
}
\end{figure}

\begin{figure}
\plotone{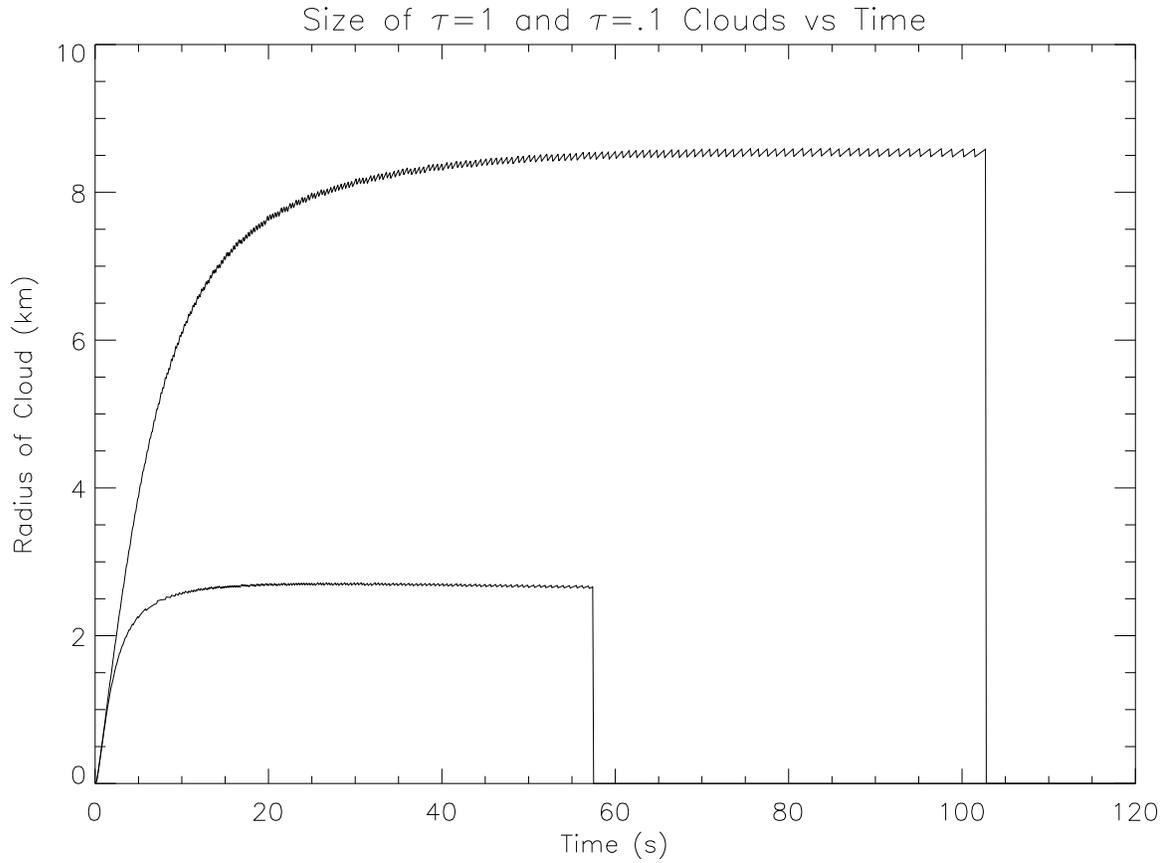}
\caption{
the time evolution of appearance in optical depth of the ``minimal TLP''
explosion, with the bottom trace showing the radius as a function of time of
the optical depth unity surface of the dust cloud, and the top trace showing
the optical depth $\tau = 0.1$, indicating the largest scale over which an
optical disturbance might be readily visible.
}
\end{figure}

\end{document}